\documentclass[iop]{emulateapj}

\shorttitle{Inclined orbit for KOI-368.01}
\shortauthors{Zhou \& Huang}

\newcommand{\teff}{\ensuremath{T_{\rm eff}}\xspace}
\newcommand{\logg}{\ensuremath{\log{g}}\xspace}
\newcommand{\vsini}{\ensuremath{v \sin{i_{\rm rot}}}\xspace}
\newcommand{\feh}{\ensuremath{\rm [Fe/H]}\xspace}
\newcommand{\chisq}{\ensuremath{\chi^2}\xspace}
\newcommand{\myemail}{george@mso.anu.edu.au}
\newcommand{\lc}{light curve}
\newcommand{\lcs}{light curves}

\newcommand{\kms}{\ensuremath{\rm km\,s^{-1}}}
\newcommand{\sqarcsec}{\ensuremath{\Box^{\prime\prime}}}

\newcommand{\pxs}{\ensuremath{\rm \arcsec pixel^{-1}}}

\usepackage{graphicx}
\usepackage{amsmath}
\usepackage{xspace}
\usepackage{subfigure}

\begin{document}

\title{A highly inclined orbit for the 110-day period M-dwarf companion KOI-368.01}

\author{George Zhou\altaffilmark{1,2} and
Chelsea X. Huang\altaffilmark{2}}

\altaffiltext{1}{Research School of Astronomy and Astrophysics, Australian National University, Cotter Rd, Weston Creek, ACT 2611, Australia; \email{\myemail}}
\altaffiltext{2}{Department of Astrophysical Sciences, 4 Ivy Lane, Peyton Hall, Princeton University, Princeton, NJ 08544}

\begin{abstract}
We report the detection of asymmetry in the transit light curves of the 110-day period companion to KOI-368, a rapidly rotating A-dwarf. The significant distortion in the transit light curve is attributed to spin-orbit misalignment between the transiting companion and the gravity darkened host star. Our analysis was based on 11 Long Cadence and 2 Short Cadence transits of KOI-368.01 from the \emph{Kepler} mission, as well as stellar parameters determined from our follow-up spectroscopic observation. We measured the true obliquity between the orbit normal and the stellar rotation axis to be $69_{-10}^{+9\,\circ}$\@. We also find a secondary eclipse event with depth $29 \pm 3\,\text{ppm}$ at phase 0.59, from which the temperature of the companion is constrained to $3060\pm50\,\text{K}$, indicating that KOI-368.01 is a late M-dwarf. The eccentricity is also calculated from the eclipse to be $0.1429\pm0.0007$\@. The long period, high obliquity, and  low eccentricity of KOI-368.01 allow us to limit a number of proposed theories for the misalignment of binary systems.
\end{abstract}

\keywords{binaries: eclipsing---stars: low-mass---techniques: photometric---stars: individual (KOI-368)}

\section{Introduction}
\label{sec:introduction}

Spin-orbit alignment is predicted for binary systems by many prominent star formation theories. Binaries formed in clouds with 
the same net rotation or with disk fragmentation are naturally 
expected to be in alignment, as confirmed in the majority of systems by indirect measurements from 
polarisation of circumstellar disks \citep[e.g.][]{Monin:1998,Wheelwright:2011}.

Surprisingly, accurate obliquities of close-in stellar binaries are rarely measured. 
For systems where both the rotation period and spectroscopic rotational line broadening can be measured,
the spin-orbit coupling can be estimated \citep{Hale:1994,Harding:2013}. Similarly, an approximate obliquity can be estimated by 
comparing the measured rotational broadening per star against 
the expected rotation rate of the spectral type \citep[e.g.][]{Weis:1974,HoweClarke:2009}. 
From these measurements, we know that spin-orbit alignment is dominant for close-in stellar binaries. 
Whilst such arguments are statically meaningful, uncertainties for individual systems 
are large due to the heavy dependence on stellar age and geometry.

Only five equal-mass \citep{Albrecht:2011,Winn:2011,Albrecht:2013b}
and three unequal-mass binaries \citep{Siverd:2012,Triaud:2013} have
obliquities precisely measured using the Rossiter-McLaughlin (RM) 
effect \citep{Rossiter:1924}. Of these measured systems, 
only the B4-B5 binary DI Herculis (period 10.6 days, eccentricity 0.489) is misaligned.
The observational limitations of RM measurements, such as the 
need for well-timed observations, mean that the sampled binary 
systems are severely biased towards the short period regime ($<40$ days).

On the other hand, the spin-orbit alignment of transiting 
planets have now been well explored.
Some 50 transiting planets have been measured to date, the majority of which also with the RM effect.
These observations have revealed the diversity of orbital
obliquities for hot-Jupiters \citep{Albrecht:2012}, 
including planets in retrograde and polar orbits 
\citep[e.g.][]{Bayliss:2010,Addison:2013}. 

Proposed explanations for the obliquity of stellar binaries are similar 
to those for planetary systems, including fragmentation of a molecular cloud \citep{Bonnell:1992, 
Bate:2010}, tilting of the protoplanetary disk \citep{Bate:2000}, 
evolution of the stellar spin axis \citep{RogersLin:2012}, or dynamical 
interactions \citep[e.g.][]{WuMurray:2003,FabryckyTremaine:2007}. 

In this study, we search for the distorted \lc\ of transits about rapidly rotating stars 
due to orbital obliquity. For a gravity darkened rapid rotator \citep{Zeipel:1924}, a companion 
with non-zero obliquity will successively block different latitudes of the stellar disk that have 
different levels of gravity darkening, resulting in a distorted and potentially asymmetric light curve 
\citep{Barnes:2009}. This is the only known method
that probes both the stellar obliquity and the projected companion orbit 
obliquity, which gives us a good estimate of the true orbit inclination. The only 
existing detections of this effect includes the planet sized companion to KOI-13, 
with a period of 1.8 days \citep{Szabo:2011,Barnes:2011}, 
and DI Her \citep{PhilippovRafikov:2013}.

KOI-368.01 was identified as a Kepler Object of Interest (KOI) in the \citet{Batalha:2013} Kepler catalog. 
The planetary candidate transits an A-type star in an 110 day period orbit. We report the 
detection of asymmetry in the transit of KOI-368.01, indicative of spin-orbit misalignment for this 
system. We also report the detection of a secondary eclipse for the system, indicating the companion is of
stellar mass.

\section{KOI-368 stellar parameters}
\label{sec:host-star-parameters}

KOI-368 (KIC 6603043) is an A-dwarf with the revised Kepler Input Catalogue stellar atmosphere 
parameters of $T_{\rm eff} = 9257\pm200\,{\rm K}$, and $\log g=4.1\pm0.3$ \citep{Pinsonneault:2012}.

To validate the stellar parameters, we obtained a high resolution spectrum of KOI-368 on 11 September 2012 using the 
ARC Echelle Spectrograph (ARCES) mounted on the Apache Point Observatory 3.5-meter telescope (APO 3.5m). ARCES is 
a cross-dispersed optical spectrograph, with slit width of 1.6$\arcsec$, giving a spectral resolution of 
$\lambda / \Delta \lambda \approx 31500$ over the wavelength region 3200--10000\AA\@. The spectrum is extracted 
and reduced using the Echelle package in \emph{IRAF}\footnote{IRAF is distributed by the National Optical 
Astronomy Observatory, which is operated by the Association of Universities for Research in Astronomy (AURA) 
under cooperative agreement with the National Science Foundation.}.

The fundamental stellar atmosphere parameters (\teff, \logg, \feh, \vsini) are derived by fitting synthetic spectra 
to the observed spectrum. A spectral library was generated with the spectral synthesis program 
SPECTRUM\footnote{\url{http://www1.appstate.edu/dept/physics/spectrum/spectrum.html}} \citep{GrayCorbally:1994}, 
using ATLAS9 model atmospheres \citep{Castelli:2004}, and the default isotopic line list provided by SPECTRUM, over 
the wavelength region 4300--6700\AA\@. For each spectral order, we perform a grid search \chisq minimisation over the 
ATLAS9 grid, centred about the KIC stellar parameters. We 1) perform an unrestricted grid search to derive a first iteration of 
stellar parameters, 2) perform the light curve fit (Section \S \ref{sec:model-fitting}), using the transit to constrain the \logg, and 
3) re-perform the spectral grid search with \logg fixed. This iterative process is often used in the characterisation of 
exoplanet host stars, and provides a more constrained \logg than possible using spectral fitting alone \citep{Sozzetti:2007}.
The final atmosphere parameters are 
$T_\text{eff} = 9200\pm200\,{\rm K}$, $\log g = 4.01 \pm 0.01$, $\text{[Fe/H]} = -0.02 \pm 0.05$, $v \sin i_\text{rot} = 79 \pm 4\,\text{km\,s}^{-1}$, $\rho = 0.221\pm0.004\,(\text{g\,cm}^{-3})$,
with errors from the scatter between orders.
We match these to the Yonsei-Yale isochrones \citep{Yi:2001}, via Monte Carlo sampling over the interpolated isochrones,
obtaining the values $M_1 = 2.3 \pm 0.1\,M_\odot$, $R_1 = 2.4 \pm 0.1\,R_\odot$. The full list of derived stellar 
properties are given in Table~\ref{tab:params}. The stellar density is later independently derived from the light curve 
in Section \S \ref{sec:model-fitting}, from which we derive a more accurate iteration of stellar parameters.

We measure the host star rotation period by identifying rotational
modulation in the Kepler Long Cadence Pre-search Data Conditioning (PDC) \lcs\ \citep{Smith:2012} 
from Q1-Q9. We masked out the primary transits and performed a Lomb-Scargle 
periodogram \citep[][]{Lomb1976} analysis
on the \lcs. A significant peak at 1.19 days was identified.
Assuming the peak is due to rotational modulation, the star 
rotates with a true velocity of $\sim100$\,\kms, and the rotation axis is inclined to our line of sight by 
$\sim 50^\circ$ (this is later fitted for in Section \S \ref{sec:model-fitting})

\section{Orbit obliquity from light curve asymmetry}
\label{sec:transit-light-curve}

\subsection{Kepler light curves}
\label{sec:light curve-model}

We make use of all available public Kepler \lcs\ for our analysis of KOI-368. These include 
11 transits (Q0-Q15, more than 1300 day) of Long Cadence (29.4min) data and 2
transits (Q8-Q9) of Short Cadence (58.84s) data.

To remove the stellar variability, we use the raw flux $(\rm SAP\_FLUX)$ obtained from the MAST archive
\footnote{http://archive.stsci.edu/kepler/data$\_$search/search.php}, with the 
out-of-transit variations corrected by the following steps from 
\citet{Huang:2013}:

a) removal of bad data points;

b) correction of systematics due to various phenomena of the spacecraft, such as safe modes and tweaks;

c) a set of cosine functions with minimum period of 1 day; 

d) a 7th order polynomial fit over the out-of-transit regions. 

The asymmetry of transit (see Section \S \ref{sec:model-fitting}) is also reproduced with 
both the raw \lcs\ and PDC \lcs. We use the \lcs\ with
above corrections to produce all the fittings below because it gives us the longest flat out-of-transit base line.     

\subsection{Gravity darkening modelling}
\label{sec:grav-model}
The gravity darkening model is generated following \citet{Barnes:2009}.
The temperature profile on the stellar surface is determined by the local 
effective gravity $g_{\rm eff}$ \citep{Zeipel:1924}. We use the passband gravity
darkening coefficient $y$ which 
directly relates the specific intensity profile of the stellar disk to 
the effective gravity profile: 
\begin{equation}
y = \left(\frac{\partial\,{\ln I(\lambda)}}{\partial\,{\ln g_{\rm eff}}}\right)_{T_{\rm eff}} \\
+ \left(\frac{{\rm d}\ln T_{\rm eff}}{{\rm d}\ln g_{\rm eff}}\right)\left(\frac{\partial\,{\ln I(\lambda)}}{\partial\,{\ln T_{\rm eff}}}\right)_{g_{\rm eff}} \,.
\end{equation}
Therefore the ratio of specific intensity $I(\lambda)$ at any position to 
the intensity at stellar pole $I_{\rm pole}(\lambda)$ can be written as 
\begin{equation}
\frac{I(\lambda)}{I_{\rm pole}(\lambda)} \propto \left(\frac{g_{\rm eff}}{g_{\rm pole}}\right)^{y} \,. 
\end{equation}
Note that $y$ relates gravity with intensity, while the commonly used exponent
$\beta_\text{gravity}$ relates gravity with effective temperature $(y \approx 4\beta_\text{gravity})$.
The effective gravity is defined as 
\begin{equation}
\vec{g}_{\rm eff} = -\Omega_{\rm grav}^2\frac{R_{\rm eq}^3}{R^2}\hat{r} \\ 
+ \Omega_{\rm rot}^2\,R_{\perp}\hat{r_{\perp}} \,.
\end{equation}
We define $\Omega_{\rm rot}$ as the stellar rotation rate 
and $\Omega_{\rm grav} = \sqrt{GMR_{\rm eq}^{-3}}$ to represent the angular 
velocity due to gravity at the equator. 
The definition of the other symbols follow the Eq.10 of \citet{Barnes:2009}, in 
which $G$ is the gravitational constant, $M$ is the mass of the star, $R$ 
and $R_{\perp}$ are the distance from the stellar center and stellar 
rotation axis to the point of question, respectively. $\hat{r}$ and 
$\hat{r_{\perp}}$ are unit vectors indicate the directions of $R$ and 
$R_{\perp}$. 
The effective gravity profile $g_{\rm eff}/g_{\rm pole}$ at the
stellar surface is then only a function of the oblateness of the star $f=(R_{\rm eq}-R_{\rm pole})/R_{\rm eq}$, 
the normalised position parameters $r$, $\theta$, and the dimensionless measure of the 
rotation rate ($w=\Omega_{\rm rot}/\Omega_{\rm grav}=v_{\rm rot}^2(R/M)$). 
Since $M$ depends proportionally to $R$ for this spectral type, the uncertainty in the gravity darkening profile depends only weakly 
on the absolute stellar mass and radius, and primarily on the stellar oblateness $f$ and the rotation velocity 
$v_{\rm rot}$. In Figure~\ref{fig:lightcurve}(a), 
we demonstrate the flux profile computed for KOI-368. If the companion crosses different latitudes
during its transits, the \lcs\ will show asymmetry depending 
on the misalignment between the companion orbit and stellar rotation axis.

\subsection{Secondary eclipse}
\label{sec:secondary-eclipse}

We search for a secondary eclipse in 
the \emph{Kepler} long cadence light curve. We find a possible eclipse event at phase 0.59 via a grid search.
The depth, inferred eccentricity,
and significance of the eclipse is characterised by the global light curve modelling described in 
Section~\ref{sec:model-fitting}. This secondary eclipse
event was independently identified by E. Agol and D. Fabrycky, as well as J. 
Coughlin and the Kepler team.  

\subsection{Fitting of system parameters}
\label{sec:model-fitting}

The transit \lcs\ are modeled using the
\citet{Nelson1972} model, implemented in an adaption of the
JKTEBOP code \citep{Proper1981,Southworth2004}. The
relevant free parameters in the transit model are orbital period $P$, transit centre $T_0$,
normalised radius sum $(R_1+R_2) / a$, radius ratio $R_2/R_1$,
line of sight inclination $i$. To allow for a secondary eclipse,
we also fit for the surface brightness ratio $S_2/S_1$, and eccentricity components $e\cos \omega$ 
and $e\sin \omega$. The quadratic limb darkening coefficients 
are fitted for in an initial minimisation routine, then held fixed for 
subsequent analyses. The final values do not deviate significantly from estimates by
\citet{Sing2010}. Jump parameters modelling gravity darkening 
include the sky-projected orbit obliquity angle $\lambda$, stellar oblateness factor $f$, 
projected stellar obliquity 
$i_\text{rot}$ (with initial value taken from Section \S \ref{sec:host-star-parameters}), 
and the \emph{Kepler} band gravity darkening exponent $y$, with initial value calculated from \citet{Claret:2011}. 
A constant flux baseline offset for each transit is removed before the global fitting.
For Kepler long cadence data, the model is 
modified by a 30 minute boxcar smooth. The best fit parameters and the 
posterior probability distribution is explored via a Markov chain Monte Carlo
(MCMC) analysis, using the \emph{emcee} MCMC ensemble sampler
\citep{ForemanMackey2012}. For each transit, we scale the flux errors such
that the reduced $\chi^2$ are at unity. This allows for errors other
than photon noise to be taken into account.

\begin{figure}
  \centering
  \begin{tabular}{c}
  \includegraphics[width=9cm]{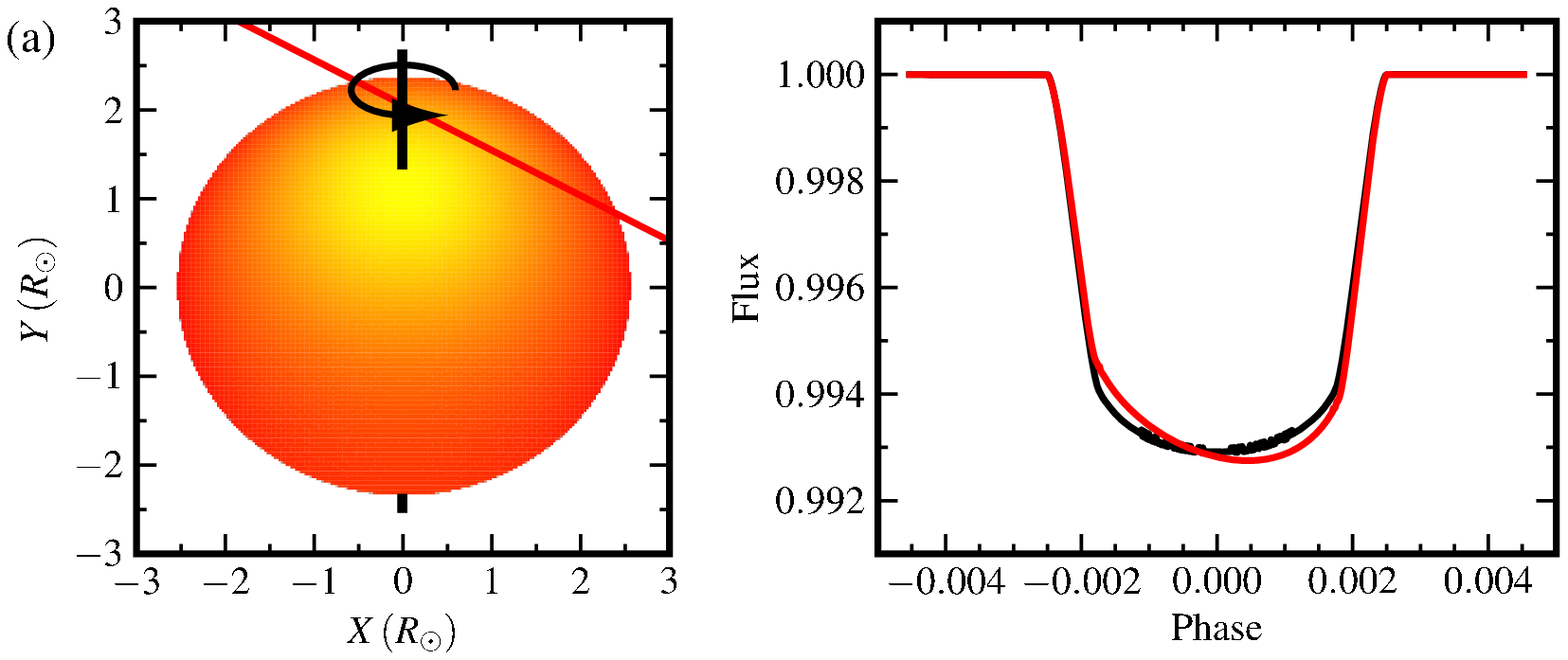}\\
  \includegraphics[width=9cm]{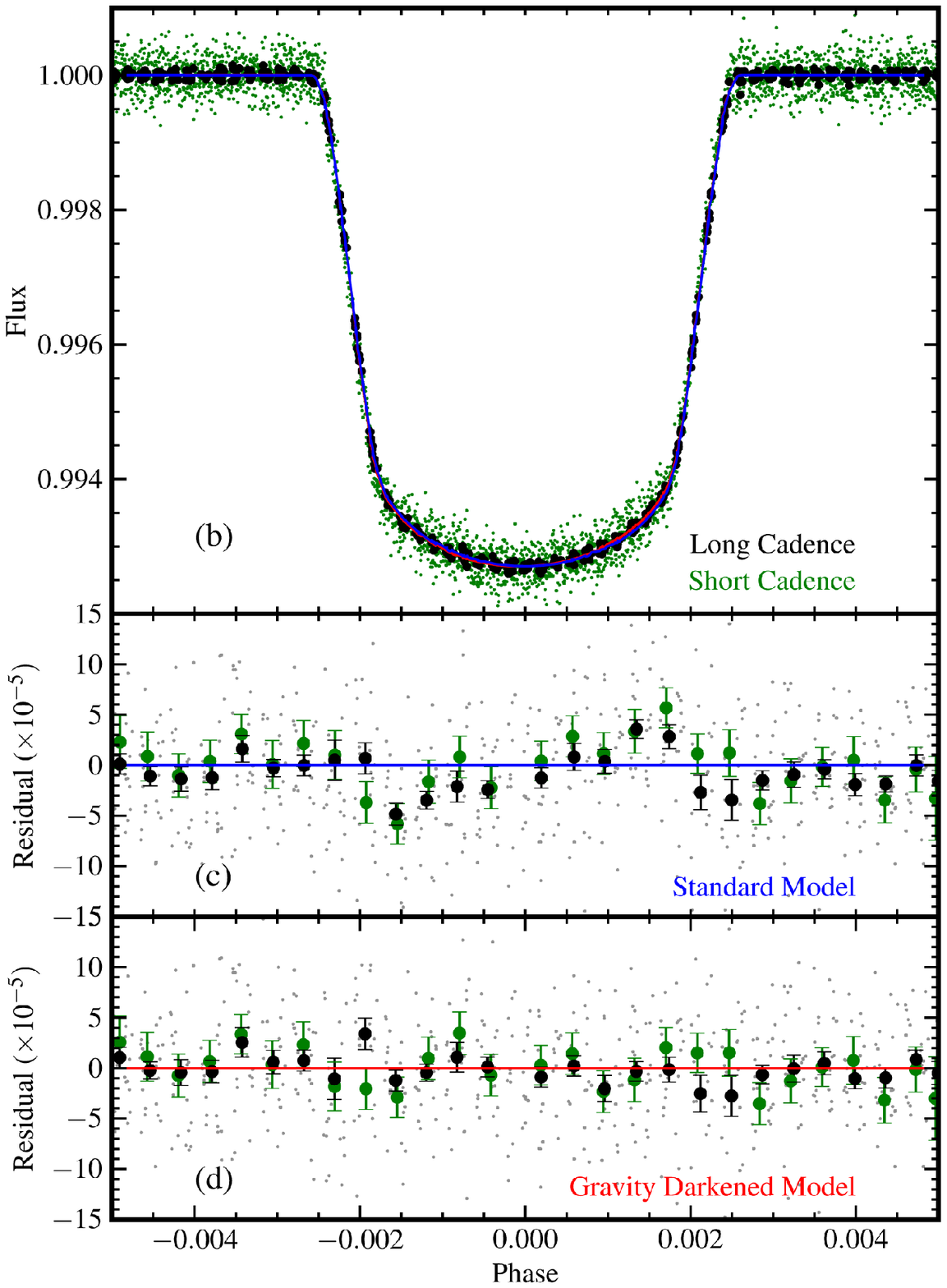}\\
  \includegraphics[width=9cm]{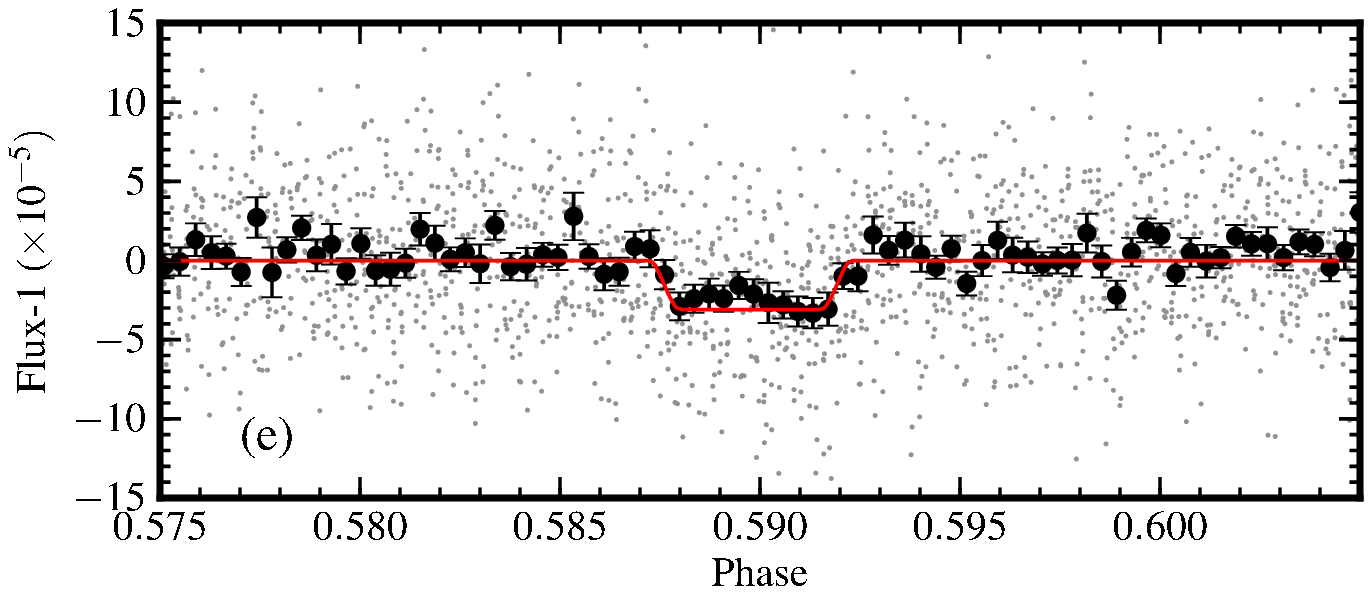}\\
  \end{tabular}

  \caption{(a) The flux ratio distribution at the surface of KOI-368 is plotted on the left, with 
  $i_\text{rot}=45^\circ$ and exaggerated oblateness parameters of $f=0.2$, $y=1.0$. An example transit path 
  with projected obliquity of $30^\circ$ is marked in red (grey). The corresponding 
  light curve is plotted on the right (in red). This is the best fit configuration
  for KOI-368.01. For comparison, the best fit light curve without 
    considering gravity darkening is also shown in black. 
  (b) Phase folded transit of KOI-368.01. Long
    cadence observations are plotted in black, short cadence in
    green. The best fit standard transit model is
    plotted in blue, gravity darkened model in red. These two models are
    visually indistinguishable at this scale, but the difference is statistically significant. (c) Residuals to the
    standard transit model. The long cadence data are
    plotted in full as gray dots, and as 1 hour binned intervals in black,
    short cadence binned residuals in green. (d)
    Residuals to the gravity darkened transit model. (e) The secondary eclipse event 
    is plotted with the best fit model.}
  \label{fig:lightcurve}
\end{figure}

Figure~\ref{fig:lightcurve} plots the phase folded transit and eclipse light curves of
KOI-368.01 and the best fit standard and gravity darkened models. The best fit
standard model cannot explain the significant in-transit
asymmetry observed. 

\begin{figure*}
  \centering
  \includegraphics[width=12cm]{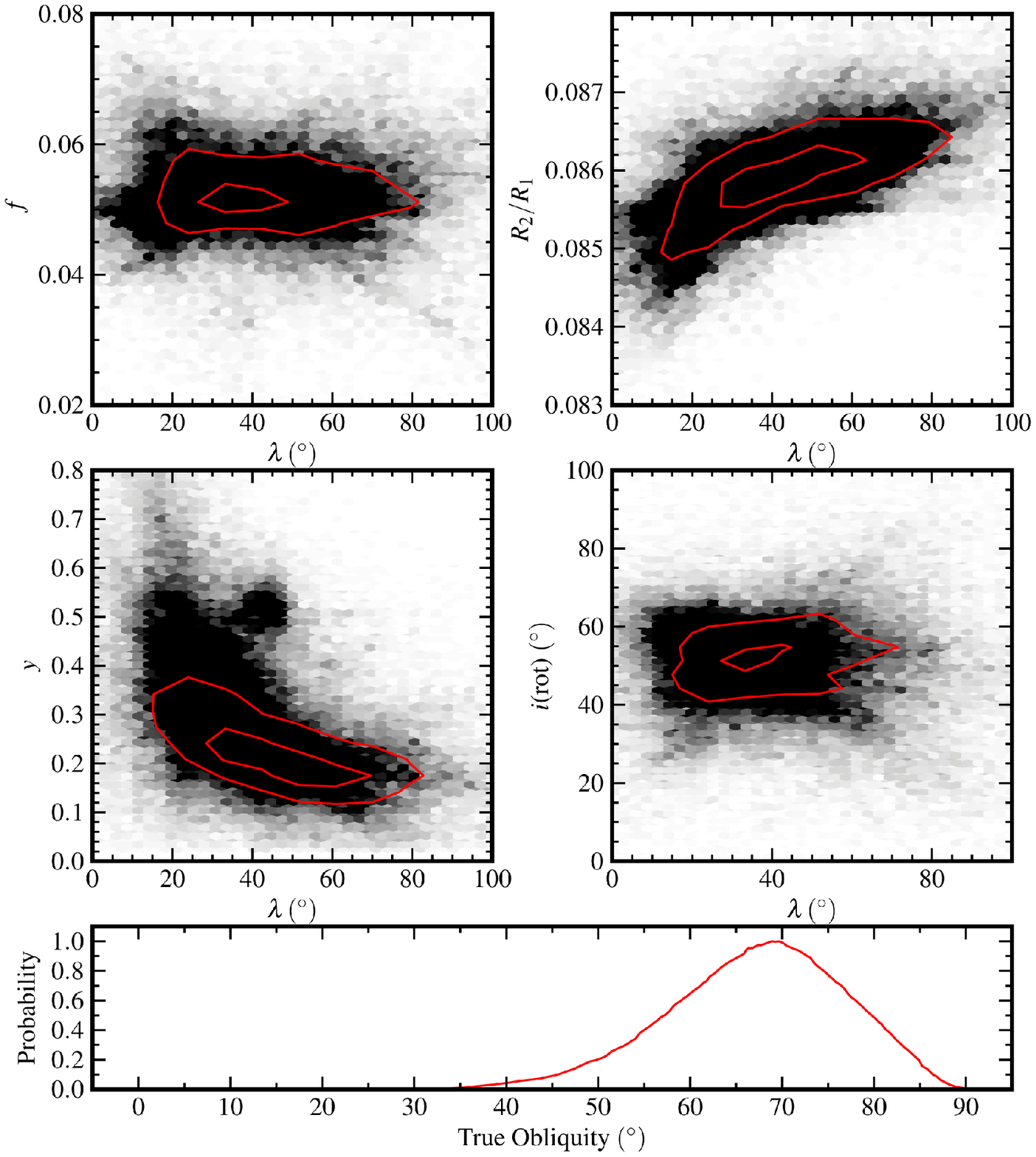}
  \caption{Posterior probability distributions showing the correlation
  between the projected obliquity $\lambda$ and stellar oblateness $f$,
  radius ratio $R_2/R_1$, gravity darkening exponent $y$, projected stellar
  obliquity $i_\text{rot}$. The
  contours mark the 1 and 2$\sigma$ confidence regions. The posterior probability distribution 
for the true obliquity is plotted on the bottom panel.}
  \label{fig:posterior}
\end{figure*}

The best fit parameters are presented in Table~\ref{tab:params}.
The posterior probability distributions for relevant parameters
are plotted in Figure~\ref{fig:posterior}. We did not detect transit
timing variations, consistent with previous studies
\citep[e.g.][]{Mazeh:2013}.

\begin{deluxetable*}{lrrr}
\tablewidth{0pc}
\tabletypesize{\scriptsize}
\tablecaption{KOI-368 System properties
\label{tab:params}}
\tablehead{
\multicolumn{1}{c}{Parameter} & 
\multicolumn{1}{c}{Kepler Catalog} &
\multicolumn{1}{c}{Standard Model} &
\multicolumn{1}{c}{Gravity Darkened Model} \\
}\startdata
\multicolumn{2}{l}{\textbf{Photometric magnitudes}}\\ 
$B$\tablenotemark{a} & 11.30 & - & - \\
$V$\tablenotemark{a} & 11.02 & - & - \\
$J$\tablenotemark{b} & 11.11 & - & - \\
$H$\tablenotemark{b} & 11.09 & - & - \\
$K$\tablenotemark{b} & 11.08 & - & - \\
$K_p$ & 11.375 & - & - \\
\\
\multicolumn{2}{l}{\textbf{Spectroscopic parameters}}\\ 
$T_\text{eff}$ (K) & $9257\pm200$ & $9200\pm200$ & -\\
$\text{[Fe/H]}$ & -0.3 \tablenotemark{c} & $-0.02\pm0.05$ & -\\
$v \sin i_\text{rot}$ (\kms) & - & $79 \pm 4$ & -\\
\\
\multicolumn{2}{l}{\textbf{Lightcurve fitting parameters \tablenotemark{b}}} \\ 
Period (Days) & $110.32160\pm{5}$ & $110.321645_{-5}^{+6}$ & $110.32164_{-1}^{+1}$ \\ 
$T_0$ $(\text{BJD}-2454000)$ & $1030.3645\pm2$ & $1030.36409_{-5}^{+3}$ & $1030.36437_{-3}^{+2}$ \\ 
$(R_2+R_1)/a$ & $0.02119\pm6$ & $0.02097_{-3}^{+5}$ & $0.0206_{-1}^{+1}$ \\ 
$R_2/R_1$ & $0.08453\pm3$ & $0.08408_{-5}^{+2}$ & $0.0863_{-4}^{+4}$ \\
$i$ & $89.38$ &$89.204_{-2}^{+5}$ & $89.235_{-7}^{+9}$ \\
$e\cos\omega$ & - &$0.1417_{-1}^{+1}$ & $0.1416_{-1}^{+1}$ \\ 
$e\sin\omega$ & - &$-0.012_{-7}^{+3}$ & $-0.019_{-5}^{+4}$ \\ 
$S_2/S_1$ & - & $0.0045 _{-2}^{+2}$ & $0.0039 _{-3}^{+3}$ \\
$f$ & - &- & $0.052_{-3}^{+3}$ \\
$y$ & - &- & $0.20_{-4}^{+6}$ \\
$\lambda$ & - &- & $36_{-17}^{+23}$ \\
$i_\text{rot}$ & - & - & $55_{-10}^{+3}$ \\
\\
\multicolumn{2}{l}{\textbf{Derived parameters}}\\ 
$\rho_1\,(\text{g\,cm}^{-3})$ \tablenotemark{e}& - & $0.216\pm0.003$ & $0.221\pm0.004$\\
$M_1\,(M_\odot)$ & - & $2.3 \pm 0.1$ & $2.3\pm0.1$\\
$R_1\,(R_\odot)$ & $2.1 \pm 0.9$ & $2.5 \pm 0.1$ & $2.4 \pm 0.1$ \\
$\log g$ & $4.1\pm0.3 $ & $4.02\pm0.01$ & $4.02\pm0.01$\\
Age (Gyr) & $0.51 \pm 0.09$ & $0.49 \pm 0.06$ & $0.48 \pm 0.06$\\
Luminosity ($L_\odot$) & - & $38\pm5$ & $38\pm6$\\
Distance (kpc) & - & $1.07\pm0.05$ & $1.07\pm0.05$ \\
$R_2\,(R_\odot)$ & $0.18\pm0.08$ & $0.209 \pm 0.007$ & $0.211 \pm 0.006$\\
True Obliquity $(^\circ)$ & - & - & $69_{-10}^{+9}$\\ 
$e$ & - & $0.142\pm0.001$ & $0.1429\pm0.0007$\\
$T_\text{2,eff}$ & - & $3080\pm50$ & $3060\pm50$\\
\enddata
\tablenotetext{a}{Tycho catalogue}
\tablenotetext{b}{2MASS catalogue}
\tablenotetext{c}{Assumed}
\tablenotetext{d}{Uncertainties quoted are for the last significant figure}
\tablenotetext{e}{Derived from light curve fit}
\end{deluxetable*}

We find a best fit projected obliquity value of $\lambda = 36_{-17}^{+23\,\circ}$, and a 
projected stellar obliquity of $i_\text{rot}=55_{-10}^{+3\, \circ}$. The derived
true spin-orbit obliquity of KOI-368.01 is $69_{-10} ^{+9\,\circ}$, indicating 
that it is orbiting in a highly inclined orbit. 

Note that due to the degeneracy between the geometries of 
$\lambda$ and $(180^\circ - \lambda)$,  we cannot distinguish between prograde and retrograde
orbits using this method. We conservatively choose the prograde solution for the quoted angles in this paper.

Using the phase and duration of the secondary eclipse, we obtain an eccentricity of
 $e = 0.1429\pm0.0007$. The expected transit duration for a circular orbit transit is 12.5 hours. The ratio between 
the observed and circular expected transit durations give an eccentricity estimate
of $e = 0.07^{+0.18}_{-0.07}$, with degeneracies constrained by the lack of 
ingress-egress duration variations. This is consistent with the eccentricity
derived from the secondary eclipse phase.

The light curve asymmetry can not be reproduced
by the in-transit velocity change of an eccentric orbit \citep{Barnes2007} 
or by the photometric RM effect \citep{Shporer:2012,Groot:2012}, both of which result
in light curve distortions at least an order of magnitude smaller than that observed.

We also fit a gravity darkening exponent for KOI-368 of $y=0.20_{-0.04}^{+0.06}$ ($\beta_\text{gravity}\approx 0.05$).
This is significantly lower than simple
theoretical model predictions, which is also seen for DI Her \citep{PhilippovRafikov:2013}.

\section{Nature of the KOI-368 system}
\label{sec:system-properties}

\subsection{Excluding blend scenarios}
\label{sec:excluding-blends}

\begin{figure}
\subfigure[]
{
\includegraphics[width=0.37\linewidth]{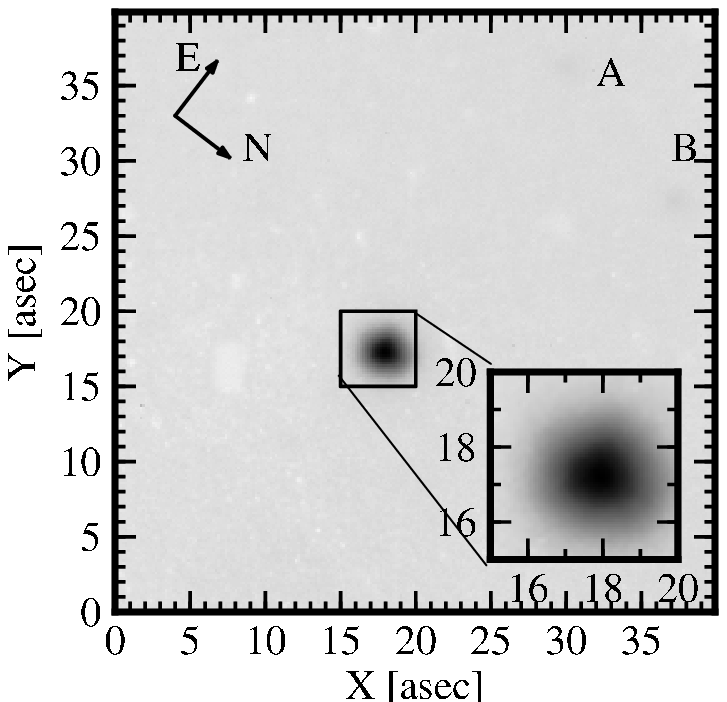}
\label{fig:APO}
}
\subfigure[]{
\includegraphics[width=0.4\linewidth]{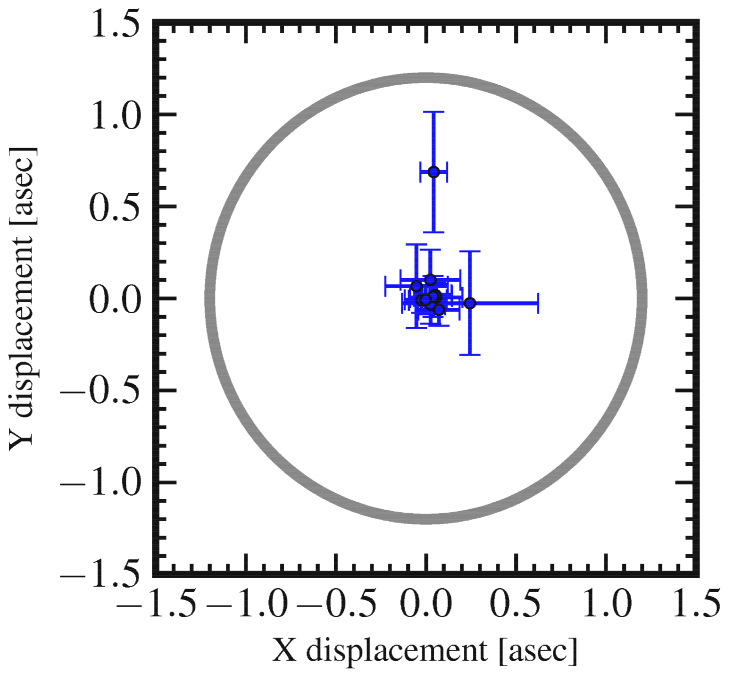}
\label{fig:centroid}
}
\caption{
Left: Image of KOI-368 from APO 3.5\,m Echelle Slitviewer 
camera, within $40\times40$\sqarcsec\ area. The two companions  
brighter than 20 magnitude (marked out as A 
and B), are $\sim 20 \arcsec$ away. Insert shows the target PSF. The half width radius of the 
PSF is 1$\arcsec$. The shape of PSF is not visibly elongated. 
Right: Centroid analysis of KOI-368. The
transit depth normalised centroids displacement between in-transit and 
out-transit for all 11 transits are shown. 
This indicates that the transit signal must be from source $>1 \arcsec$ 
(the black circle). 
\label{fig:FP}
}
\end{figure}

To constrain the possible blend scenarios for KOI-368, we obtained an image 
of the object using the APO 3.5m 
Echelle Slitviewer camera. The 
camera has a field of view of 63.6$\arcsec$ and pixel scale of 0.133\pxs. The 
full width half maximum of the point spread function is 1$\arcsec$. 
The closest companion brighter than 20 mag is more than 20$\arcsec$ away from KOI-368 (See Figure \ref{fig:APO}).
We can exclude blend sources with separations greater than $\sim 1\arcsec$ from the object.

We can also constrain the possibility of blends by searching for centroid displacements
in and out of transit. The displacements are given by 
the flux weighted first momentum centroids produced by Kepler 
pipeline. We correct for the low frequency ($>5$ day) trends
in the centroids with a 7th order polynomial around each transit.
The displacements between the mean centroids in- and  
out-of-transit, weighted by the transit depth 
\citep{Chaplin:2013}, are shown in Figure \ref{fig:centroid}. The uncertainties 
are computed assuming a Gaussian distribution for centroids in each transit.
The lack of displacement allows us to rule
out an eclipsing neighbour $> 1\arcsec$ away.

\subsection{KOI-368.01 is an M-dwarf}

We can constrain the secondary eclipse to be of depth $29 \pm 3\,\text{ppm}$. 
Using spectral models from \citep{Gustafsson:2008}, and integrating over the Kepler band,
we find the temperature of the companion to be $3060\pm60\,\text{K}$. 
Given that the derived radius of the companion is $0.211\pm0.006\,\text{R}_\odot$, the
companion is most consistent with an M-dwarf.

\section{Discussion}
\label{sec:discussion}

We characterised the asymmetric transit and eclipse light curve of KOI-368.01, and found it to be an M-dwarf
companion orbiting in a highly inclined $(69_{-10} ^{+9\,\circ})$, near-circular $(e = 0.1429\pm0.0007)$ orbit about a rapidly
rotating A-dwarf. 

\begin{figure}
  \centering
  \includegraphics[width=9cm]{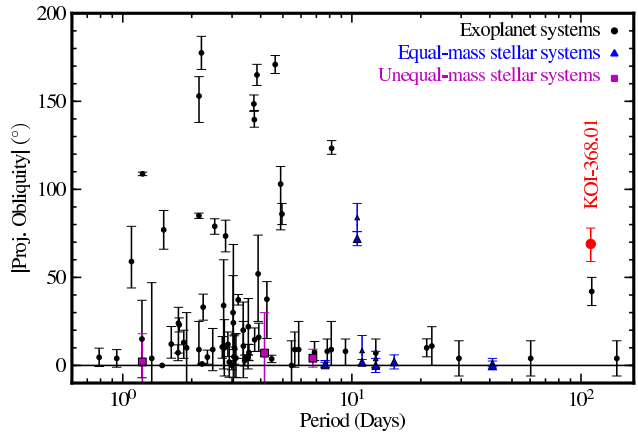}
  \caption{The obliquities of binary systems are plotted against their orbital period. Equal-mass (same spectral type components) stellar systems are marked by blue triangles, with the obliquity of the primary and secondary differentiated by the symbol size \citep{Albrecht:2011,Winn:2011,Harding:2013,Albrecht:2013b}. Unequal-mass system host stars are labelled by magenta squares \citep{Siverd:2012,Triaud:2013}. Exoplanet hosts are plotted in black.}
  \label{fig:periodobliq}
\end{figure}

KOI-368 is a distinct system in the period-obliquity space, in the context of both binary and planetary systems 
(Figure~\ref{fig:periodobliq}). 
With a period of 110.3 days, KOI-368 is one of the longest 
period systems for which spin-orbit alignment has been precisely 
measured. Other measured highly misaligned systems mostly have 
shorter period orbits, such as DI Her (10.55 days). 
Curiously, the orbit of KOI-368.01 is almost 
identical with the planet HD80606b (111.4 day, $e=0.93$) in both period and 
projected orbit obliquity, but with a much lower eccentricity. 
HD80606b can be explained 
satisfactorily with Kozai cycle excited by its main-sequence 
companion 1000 AU away \citep{WuMurray:2003, FabryckyTremaine:2007}.  
For KOI-368.01 to be in a high eccentricity 
migration track, one would expect $e\gtrsim\,0.8$, such that the 
final semi-major axis would be $a<0.1{\rm AU}$. 
The age of the system is ~500 Myrs, whilst the circularisation timescale of this system
is much longer than the Hubble time. Alternatively, the high obliquity of this system 
can be excited by a third body in this system with Kozai mechanism 
\citep[e.g.][]{FabryckyTremaine:2007,Naoz:2011,Katz:2011} after migrated to the current location.
More complex dynamic interactions, such as sequential Kozai migration, can also result in 
the current configuration \citep[e.g.][]{Narita:2012}.

While the obliquity of KOI-368 is also possibly primordial, 
we can limit the main scenarios for stellar binary misalignment.
Simple rotational fragmentation of molecular clouds results in 
aligned systems. \citet{Bonnell:1992} proposed that rotational fragmentation of a
highly elongated, tilted, molecular cloud can form misaligned binaries. However, 
the systems produced by this mechanism tend to have a high initial 
eccentricity (0.4-0.9). Chaotic accretion has also been suggested 
as a mechanism for tilting the circumstellar disk \citep{Bate:2010}, but
results in a misaligned low-mass disk that is unlikely to form stellar mass 
companions.

We highlight the advantages of this technique in measuring spin-orbit alignments 
for long-period systems. Ground based RM observations require careful planning, fortunate weather, and are limited
in transit duration. Observations of KOI-368 are also limited by its high \teff and large \vsini.
Measuring spin-orbit alignment via star-spot crossings, the only other
technique to date to yield results for 
such long-period systems, is severely biased towards binaries in perfect spin-orbit alignment, 
and will not identify systems like KOI-368. The high-precision photometry from \emph{Kepler} means that 
measurements of orbit obliquity from companions about rapidly rotating stars, in conjunction with 
measurements of the stellar obliquity, are crucial in exploring the dynamics 
of long-period stellar binaries and planetary systems.

\acknowledgments
The authors thank the reviewers for their insightful suggestions, G.\'A. Bakos for his comments on the letter draft, J.D. Hartman and C. Petrovich for helpful discussions, W.A. Bhatti, J.L. Prieto, and S. Dong for the APO 3.5\,m observations. We thank independent comments by E. Agol \& D. Fabricky and J. Coughlin \& the Kepler team for bringing to our attention the secondary eclipse event; A. Shporer for the photometric RM effect. Work by XCH is supported by the NASA NNX12AH91H and NSF AST1108686 grants. Work by GZ is supported by the NASA NNX12AH91H grant and the Princeton University VSRC program.

\bibliographystyle{apj}

\begin{thebibliography}{51}
\expandafter\ifx\csname natexlab\endcsname\relax\def\natexlab#1{#1}\fi

\bibitem[{{Addison} {et~al.}(2013){Addison}, {Tinney}, {Wright}, {Bayliss},
  {Zhou}, {Hartman}, {Bakos}, \& {Schmidt}}]{Addison:2013}
{Addison}, B.~C., {Tinney}, C.~G., {Wright}, D.~J., {et~al.} 2013, ArXiv
  e-prints

\bibitem[{{Albrecht} {et~al.}(2013){Albrecht}, {Setiawan}, {Torres},
  {Fabrycky}, \& {Winn}}]{Albrecht:2013b}
{Albrecht}, S., {Setiawan}, J., {Torres}, G., {Fabrycky}, D.~C., \& {Winn},
  J.~N. 2013, \apj, 767, 32

\bibitem[{{Albrecht} {et~al.}(2011){Albrecht}, {Winn}, {Carter}, {Snellen}, \&
  {de Mooij}}]{Albrecht:2011}
{Albrecht}, S., {Winn}, J.~N., {Carter}, J.~A., {Snellen}, I.~A.~G., \& {de
  Mooij}, E.~J.~W. 2011, \apj, 726, 68

\bibitem[{{Albrecht} {et~al.}(2012){Albrecht}, {Winn}, {Johnson}, {Howard},
  {Marcy}, {Butler}, {Arriagada}, {Crane}, {Shectman}, {Thompson}, {Hirano},
  {Bakos}, \& {Hartman}}]{Albrecht:2012}
{Albrecht}, S., {Winn}, J.~N., {Johnson}, J.~A., {et~al.} 2012, \apj, 757, 18

\bibitem[{{Barnes}(2007)}]{Barnes2007}
{Barnes}, J.~W. 2007, \pasp, 119, 986

\bibitem[{{Barnes}(2009)}]{Barnes:2009}
---. 2009, \apj, 705, 683

\bibitem[{{Barnes} {et~al.}(2011){Barnes}, {Linscott}, \&
  {Shporer}}]{Barnes:2011}
{Barnes}, J.~W., {Linscott}, E., \& {Shporer}, A. 2011, \apjs, 197, 10

\bibitem[{{Batalha} {et~al.}(2013){Batalha}, {Rowe}, {Bryson}, {Barclay},
  {Burke}, {Caldwell}, {Christiansen}, {Mullally}, {Thompson}, {Brown},
  {Dupree}, {Fabrycky}, {Ford}, {Fortney}, {Gilliland}, {Isaacson}, {Latham},
  {Marcy}, {Quinn}, {Ragozzine}, {Shporer}, {Borucki}, {Ciardi}, {Gautier},
  {Haas}, {Jenkins}, {Koch}, {Lissauer}, {Rapin}, {Basri}, {Boss}, {Buchhave},
  {Carter}, {Charbonneau}, {Christensen-Dalsgaard}, {Clarke}, {Cochran},
  {Demory}, {Desert}, {Devore}, {Doyle}, {Esquerdo}, {Everett}, {Fressin},
  {Geary}, {Girouard}, {Gould}, {Hall}, {Holman}, {Howard}, {Howell},
  {Ibrahim}, {Kinemuchi}, {Kjeldsen}, {Klaus}, {Li}, {Lucas}, {Meibom},
  {Morris}, {Pr{\v s}a}, {Quintana}, {Sanderfer}, {Sasselov}, {Seader},
  {Smith}, {Steffen}, {Still}, {Stumpe}, {Tarter}, {Tenenbaum}, {Torres},
  {Twicken}, {Uddin}, {Van Cleve}, {Walkowicz}, \& {Welsh}}]{Batalha:2013}
{Batalha}, N.~M., {Rowe}, J.~F., {Bryson}, S.~T., {et~al.} 2013, \apjs, 204, 24

\bibitem[{{Bate} {et~al.}(2000){Bate}, {Bonnell}, {Clarke}, {Lubow}, {Ogilvie},
  {Pringle}, \& {Tout}}]{Bate:2000}
{Bate}, M.~R., {Bonnell}, I.~A., {Clarke}, C.~J., {et~al.} 2000, \mnras, 317,
  773

\bibitem[{{Bate} {et~al.}(2010){Bate}, {Lodato}, \& {Pringle}}]{Bate:2010}
{Bate}, M.~R., {Lodato}, G., \& {Pringle}, J.~E. 2010, \mnras, 401, 1505

\bibitem[{{Bayliss} {et~al.}(2010){Bayliss}, {Winn}, {Mardling}, \&
  {Sackett}}]{Bayliss:2010}
{Bayliss}, D.~D.~R., {Winn}, J.~N., {Mardling}, R.~A., \& {Sackett}, P.~D.
  2010, \apjl, 722, L224

\bibitem[{{Bonnell} {et~al.}(1992){Bonnell}, {Arcoragi}, {Martel}, \&
  {Bastien}}]{Bonnell:1992}
{Bonnell}, I., {Arcoragi}, J.-P., {Martel}, H., \& {Bastien}, P. 1992, \apj,
  400, 579

\bibitem[{{Castelli} \& {Kurucz}(2004)}]{Castelli:2004}
{Castelli}, F., \& {Kurucz}, R.~L. 2004, ArXiv Astrophysics e-prints

\bibitem[{{Chaplin} {et~al.}(2013){Chaplin}, {Sanchis-Ojeda}, {Campante},
  {Handberg}, {Stello}, {Winn}, {Basu}, {Christensen-Dalsgaard}, {Davies},
  {Metcalfe}, {Buchhave}, {Fischer}, {Bedding}, {Cochran}, {Elsworth},
  {Gilliland}, {Hekker}, {Huber}, {Isaacson}, {Karoff}, {Kawaler}, {Kjeldsen},
  {Latham}, {Lund}, {Lundkvist}, {Marcy}, {Miglio}, {Barclay}, \&
  {Lissauer}}]{Chaplin:2013}
{Chaplin}, W.~J., {Sanchis-Ojeda}, R., {Campante}, T.~L., {et~al.} 2013, \apj,
  766, 101

\bibitem[{{Claret} \& {Bloemen}(2011)}]{Claret:2011}
{Claret}, A., \& {Bloemen}, S. 2011, \aap, 529, A75

\bibitem[{{Fabrycky} \& {Tremaine}(2007)}]{FabryckyTremaine:2007}
{Fabrycky}, D., \& {Tremaine}, S. 2007, \apj, 669, 1298

\bibitem[{{Foreman-Mackey} {et~al.}(2012){Foreman-Mackey}, {Hogg}, {Lang}, \&
  {Goodman}}]{ForemanMackey2012}
{Foreman-Mackey}, D., {Hogg}, D.~W., {Lang}, D., \& {Goodman}, J. 2012, ArXiv
  e-prints

\bibitem[{{Gray} \& {Corbally}(1994)}]{GrayCorbally:1994}
{Gray}, R.~O., \& {Corbally}, C.~J. 1994, \aj, 107, 742

\bibitem[{{Groot}(2012)}]{Groot:2012}
{Groot}, P.~J. 2012, \apj, 745, 55

\bibitem[{{Gustafsson} {et~al.}(2008){Gustafsson}, {Edvardsson}, {Eriksson},
  {J{\o}rgensen}, {Nordlund}, \& {Plez}}]{Gustafsson:2008}
{Gustafsson}, B., {Edvardsson}, B., {Eriksson}, K., {et~al.} 2008, \aap, 486,
  951

\bibitem[{{Hale}(1994)}]{Hale:1994}
{Hale}, A. 1994, \aj, 107, 306

\bibitem[{{Harding} {et~al.}(2013){Harding}, {Hallinan}, {Konopacky},
  {Kratter}, {Boyle}, {Butler}, \& {Golden}}]{Harding:2013}
{Harding}, L.~K., {Hallinan}, G., {Konopacky}, Q.~M., {et~al.} 2013, \aap, 554,
  A113

\bibitem[{{Howe} \& {Clarke}(2009)}]{HoweClarke:2009}
{Howe}, K.~S., \& {Clarke}, C.~J. 2009, \mnras, 392, 448

\bibitem[{{Huang} {et~al.}(2013){Huang}, {Bakos}, \& {Hartman}}]{Huang:2013}
{Huang}, X., {Bakos}, G.~{\'A}., \& {Hartman}, J.~D. 2013, \mnras, 429, 2001

\bibitem[{{Katz} {et~al.}(2011){Katz}, {Dong}, \& {Malhotra}}]{Katz:2011}
{Katz}, B., {Dong}, S., \& {Malhotra}, R. 2011, Physical Review Letters, 107,
  181101

\bibitem[{{Lomb}(1976)}]{Lomb1976}
{Lomb}, N.~R. 1976, \apss, 39, 447

\bibitem[{{Mazeh} {et~al.}(2013){Mazeh}, {Nachmani}, {Holczer}, {Fabrycky},
  {Ford}, {Sanchis-Ojeda}, {Sokol}, {Rowe}, {Zucker}, {Agol}, {Carter},
  {Lissauer}, {Quintana}, {Ragozzine}, {Steffen}, \& {Welsh}}]{Mazeh:2013}
{Mazeh}, T., {Nachmani}, G., {Holczer}, T., {et~al.} 2013, ArXiv e-prints

\bibitem[{{Monin} {et~al.}(1998){Monin}, {Menard}, \& {Duchene}}]{Monin:1998}
{Monin}, J.-L., {Menard}, F., \& {Duchene}, G. 1998, \aap, 339, 113

\bibitem[{{Naoz} {et~al.}(2011){Naoz}, {Farr}, {Lithwick}, {Rasio}, \&
  {Teyssandier}}]{Naoz:2011}
{Naoz}, S., {Farr}, W.~M., {Lithwick}, Y., {Rasio}, F.~A., \& {Teyssandier}, J.
  2011, \nat, 473, 187

\bibitem[{{Narita} {et~al.}(2012){Narita}, {Takahashi}, {Kuzuhara}, {Hirano},
  {Suenaga}, {Kandori}, {Kudo}, {Sato}, {Suzuki}, {Ida}, {Nagasawa}, {Abe},
  {Brandner}, {Brandt}, {Carson}, {Egner}, {Feldt}, {Goto}, {Grady}, {Guyon},
  {Hashimoto}, {Hayano}, {Hayashi}, {Hayashi}, {Henning}, {Hodapp}, {Ishii},
  {Iye}, {Janson}, {Knapp}, {Kusakabe}, {Kwon}, {Matsuo}, {Mayama}, {McElwain},
  {Miyama}, {Morino}, {Moro-Martin}, {Nishimura}, {Pyo}, {Serabyn}, {Suto},
  {Takami}, {Takato}, {Terada}, {Thalmann}, {Tomono}, {Turner}, {Watanabe},
  {Wisniewski}, {Yamada}, {Takami}, {Usuda}, \& {Tamura}}]{Narita:2012}
{Narita}, N., {Takahashi}, Y.~H., {Kuzuhara}, M., {et~al.} 2012, \pasj, 64, L7

\bibitem[{{Nelson} \& {Davis}(1972)}]{Nelson1972}
{Nelson}, B., \& {Davis}, W.~D. 1972, \apj, 174, 617

\bibitem[{{Philippov} \& {Rafikov}(2013)}]{PhilippovRafikov:2013}
{Philippov}, A.~A., \& {Rafikov}, R.~R. 2013, \apj, 768, 112

\bibitem[{{Pinsonneault} {et~al.}(2012){Pinsonneault}, {An},
  {Molenda-{\.Z}akowicz}, {Chaplin}, {Metcalfe}, \&
  {Bruntt}}]{Pinsonneault:2012}
{Pinsonneault}, M.~H., {An}, D., {Molenda-{\.Z}akowicz}, J., {et~al.} 2012,
  \apjs, 199, 30

\bibitem[{{Popper} \& {Etzel}(1981)}]{Proper1981}
{Popper}, D.~M., \& {Etzel}, P.~B. 1981, \aj, 86, 102

\bibitem[{{Rogers} {et~al.}(2012){Rogers}, {Lin}, \& {Lau}}]{RogersLin:2012}
{Rogers}, T.~M., {Lin}, D.~N.~C., \& {Lau}, H.~H.~B. 2012, \apjl, 758, L6

\bibitem[{{Rossiter}(1924)}]{Rossiter:1924}
{Rossiter}, R.~A. 1924, \apj, 60, 15

\bibitem[{{Shporer} {et~al.}(2012){Shporer}, {Brown}, {Mazeh}, \&
  {Zucker}}]{Shporer:2012}
{Shporer}, A., {Brown}, T., {Mazeh}, T., \& {Zucker}, S. 2012, New Astronomy,
  17, 309

\bibitem[{{Sing}(2010)}]{Sing2010}
{Sing}, D.~K. 2010, \aap, 510, A21

\bibitem[{{Siverd} {et~al.}(2012){Siverd}, {Beatty}, {Pepper}, {Eastman},
  {Collins}, {Bieryla}, {Latham}, {Buchhave}, {Jensen}, {Crepp}, {Street},
  {Stassun}, {Gaudi}, {Berlind}, {Calkins}, {DePoy}, {Esquerdo}, {Fulton},
  {F{\H u}r{\'e}sz}, {Geary}, {Gould}, {Hebb}, {Kielkopf}, {Marshall}, {Pogge},
  {Stanek}, {Stefanik}, {Szentgyorgyi}, {Trueblood}, {Trueblood}, {Stutz}, \&
  {van Saders}}]{Siverd:2012}
{Siverd}, R.~J., {Beatty}, T.~G., {Pepper}, J., {et~al.} 2012, \apj, 761, 123

\bibitem[{{Smith} {et~al.}(2012){Smith}, {Stumpe}, {Van Cleve}, {Jenkins},
  {Barclay}, {Fanelli}, {Girouard}, {Kolodziejczak}, {McCauliff}, {Morris}, \&
  {Twicken}}]{Smith:2012}
{Smith}, J.~C., {Stumpe}, M.~C., {Van Cleve}, J.~E., {et~al.} 2012, \pasp, 124,
  1000

\bibitem[{{Southworth} {et~al.}(2004){Southworth}, {Maxted}, \&
  {Smalley}}]{Southworth2004}
{Southworth}, J., {Maxted}, P.~F.~L., \& {Smalley}, B. 2004, \mnras, 351, 1277

\bibitem[{{Sozzetti} {et~al.}(2007){Sozzetti}, {Torres}, {Charbonneau},
  {Latham}, {Holman}, {Winn}, {Laird}, \& {O'Donovan}}]{Sozzetti:2007}
{Sozzetti}, A., {Torres}, G., {Charbonneau}, D., {et~al.} 2007, \apj, 664, 1190

\bibitem[{{Szab{\'o}} {et~al.}(2011){Szab{\'o}}, {Szab{\'o}}, {Benk{\H o}},
  {Lehmann}, {Mez{\H o}}, {Simon}, {K{\H o}v{\'a}ri}, {Hodos{\'a}n},
  {Reg{\'a}ly}, \& {Kiss}}]{Szabo:2011}
{Szab{\'o}}, G.~M., {Szab{\'o}}, R., {Benk{\H o}}, J.~M., {et~al.} 2011, \apjl,
  736, L4

\bibitem[{{Triaud} {et~al.}(2013){Triaud}, {Hebb}, {Anderson}, {Cargile},
  {Collier Cameron}, {Doyle}, {Faedi}, {Gillon}, {Gomez Maqueo Chew},
  {Hellier}, {Jehin}, {Maxted}, {Naef}, {Pepe}, {Pollacco}, {Queloz},
  {S{\'e}gransan}, {Smalley}, {Stassun}, {Udry}, \& {West}}]{Triaud:2013}
{Triaud}, A.~H.~M.~J., {Hebb}, L., {Anderson}, D.~R., {et~al.} 2013, \aap, 549,
  A18

\bibitem[{{von Zeipel}(1924)}]{Zeipel:1924}
{von Zeipel}, H. 1924, \mnras, 84, 684

\bibitem[{{Weis}(1974)}]{Weis:1974}
{Weis}, E.~W. 1974, \apj, 190, 331

\bibitem[{{Wheelwright} {et~al.}(2011){Wheelwright}, {Vink}, {Oudmaijer}, \&
  {Drew}}]{Wheelwright:2011}
{Wheelwright}, H.~E., {Vink}, J.~S., {Oudmaijer}, R.~D., \& {Drew}, J.~E. 2011,
  \aap, 532, A28

\bibitem[{{Winn} {et~al.}(2011){Winn}, {Albrecht}, {Johnson}, {Torres},
  {Cochran}, {Marcy}, {Howard}, {Isaacson}, {Fischer}, {Doyle}, {Welsh},
  {Carter}, {Fabrycky}, {Ragozzine}, {Quinn}, {Shporer}, {Howell}, {Latham},
  {Orosz}, {Prsa}, {Slawson}, {Borucki}, {Koch}, {Barclay}, {Boss},
  {Christensen-Dalsgaard}, {Girouard}, {Jenkins}, {Klaus}, {Meibom}, {Morris},
  {Sasselov}, {Still}, \& {Van Cleve}}]{Winn:2011}
{Winn}, J.~N., {Albrecht}, S., {Johnson}, J.~A., {et~al.} 2011, \apjl, 741, L1

\bibitem[{{Wu} \& {Murray}(2003)}]{WuMurray:2003}
{Wu}, Y., \& {Murray}, N. 2003, \apj, 589, 605

\bibitem[{{Yi} {et~al.}(2001){Yi}, {Demarque}, {Kim}, {Lee}, {Ree}, {Lejeune},
  \& {Barnes}}]{Yi:2001}
{Yi}, S., {Demarque}, P., {Kim}, Y.-C., {et~al.} 2001, \apjs, 136, 417

\end{thebibliography}

\end{document}